\documentclass[a4paper,11pt]{article}
\usepackage{pos}
\usepackage{caption}
\usepackage{lipsum}
\usepackage{hyperref}
\usepackage{wrapfig}  
\usepackage{lineno}
\usepackage{graphicx}
\usepackage{subcaption}
\title{Machine learning driven reconstruction of cosmic-ray air showers for next generation radio arrays}
\ShortTitle{Cosmic-ray radio reconstruction using machine learning}

\author{The IceCube-Gen2 Collaboration \\{\normalsize \normalfont(a complete list of authors can be found at the end of the proceedings)}\\}
\emailAdd{paras.koundal@icecube.wisc.edu}

\abstract{Surface radio antenna-based measurements of cosmic-ray air showers present significant computational challenges in accurately reconstructing physics observables, in particular, the depth of shower maximum, X$_{max}$. State-of-the-art template fitting methods rely on extensive simulation libraries, limiting scalability. This work introduces a technique utilizing graph neural networks to reconstruct key air-shower parameters, in particular, direction and shower-core, energy, and X$_{max}$. For training and testing of the networks, we use a CoREAS simulation library made for a future enhancement of IceCube’s surface array with radio antennas. The neural networks provide a scalable framework for large-scale data analysis for next-generation astroparticle observatories, such as IceCube-Gen2.

\vspace{4mm}
{\bfseries Corresponding authors:}
Paras Koundal$^{1*}$,\\
{$^{1}$ \itshape Bartol Research Institute, Department of Physics and Astronomy, University of Delaware, U.S.A.}\\[4mm]
$^*$ Presenter
}

\ConferenceLogo{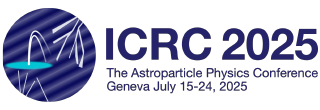}

\FullConference{39th International Cosmic Ray Conference (ICRC2025)\\
 15–24 July 2025\\
Geneva, Switzerland\\}

\newcommand{\Xmax}{$X_\mathrm{max}~$}
\newcommand{\ChiSq}{$\chi^{2}$}
\newcommand{\vxb}{$\vec{v}\times\vec{B}~$}
\newcommand{\vxvxb}{$\vec{v}\times(\vec{v}\times\vec{B})~$}
\begin{document}
\maketitle

\section{Introduction}\label{sec:intro}
Cosmic-ray (CR) primaries in atmosphere initiate a cascade of particles, generally referred to as an extensive air shower (EAS). The propagation of cascade particles (primarily e$^{+}$ and e$^{-}$) through Earth's atmosphere is responsible for radio emission. This emission can be measured in MHz frequency band using radio-antennas at detectors like the upcoming surface enhancement of IceCube Observatory \cite{ShefaliICRC2025, MeghaICRC2025}. The radio footprint carries information about CR primary direction, energy, and primary type \cite{SCHRODER20171}. Radio-detection allows for almost a 100$\%$ duty cycle, and additionally, antennas are much cheaper to deploy than other kinds of cosmic-ray detectors. Because of significant improvements in understanding of radio-emission mechanisms, their measurement, and digital signal processing techniques, the method has gained prominence in the last two decades \cite{Huege2016}. Radio-detection of EASs also allows an estimation of \Xmax, i.e., the atmospheric depth at which we expect the maxima in number of particles with lesser dependence on the choice of hadronic model. \Xmax is one of the most useful EAS observables to discriminate between the nuclei initiating the EASs \cite{Flaggs2024}.\par 
 Current radio-based observatories use surface detectors such as ice or water-based Cherenkov detectors and scintillators to provide the trigger for the radio antennas. The signal from the various EAS components at these surface detectors is then used to get an estimate of core-location, direction (zenith, azimuth), and energy. Radio detectors are then utilized to estimate \Xmax. Current state-of-the-art reconstruction methods rely on a template fitting approach by creating a library of simulation of simulations for each detected event (details upcoming in \autoref{sec:method}). These methods hence have reliance on other detectors for some of the observables reconstruction, and are generally computationally expensive for \Xmax estimation. The computational-cost limits applicability to the CR observatories like the surface enhancement of IceCube, Pierre-Auger Observatory, SKA, and others. This work hence focuses on developing a pipeline utilizing both physics-motivated and Graph Neural Network (GNN) based reconstructions to provide a fast and computationally efficient solution, relying solely on measurements done using radio antennas. Section \ref{sec:method} provides a concise overview of the current state-of-the-art method, describes the simulations used for GNN training, mentions the applied quality cuts, and breakdowns the reconstruction pipeline shown in \autoref{fig:Pipeline}. Section \ref{sec:performance} presents the reconstruction performance for direction, core-location, energy, and \Xmax, followed up by the summary and outlook in Section \ref{sec:outllok}.
\section{Methodology}\label{sec:method}
Current EAS reconstruction for radio detectors uses other detectors to estimate the core position, arrival direction, and energy, which are then used as seeds in simulations that vary only \Xmax. This allows for performing a \ChiSq-minimization analysis. This is done by antenna-wise comparison of the signal expectation at each location in the simulation with the observed signal. An aggregated \ChiSq~is obtained by combining all the \ChiSq~ from antenna-wise comparison to get a single quantifier for each data-simulation pair \cite{LofarXmax}. This step is repeated for each of the simulations produced in the library for that particular data-event. The value corresponding to the minima in  \ChiSq-values obtained by comparison of each data-simulation pair is then utilized to assign the estimated \Xmax for the specific event. The \Xmax resolution obtained from such methods is around 15-50 g$\cdot$cm$^{-2}$ \cite{AbdulHalim2024}. It is evident that the currently used method not only relies on other detectors to provide an estimate of the core, direction, and energy, but also relies on the creation of a library of simulations for the estimation of \Xmax for each detected event. This creates dependencies and a computational bottleneck for any analysis.\par
The \ChiSq-method discussed earlier to estimate \Xmax is effectively a pattern matching method; we are trying to estimate which data-simulation pair looks most similar to each other. The central premise of this analysis is that the pattern-matching task inherent in the \ChiSq-analysis can be reformulated as a pattern recognition problem, which can be effectively addressed using deep neural networks (DNNs). Hence, instead of generating specific simulations for each data event to perform pattern matching, DNNs can be utilized by producing a library of simulations (covering the phase-space of all observable ranges) and training only once. This also allows fast estimates of various observables at inference time. Additionally, the pattern recognition task using DNNs can also be applied not only to estimate \Xmax, but also to estimate the core, direction, and energy. 
\begin{wrapfigure}{r}{0.5\textwidth}
    \centering
    \includegraphics[width=.9\linewidth]{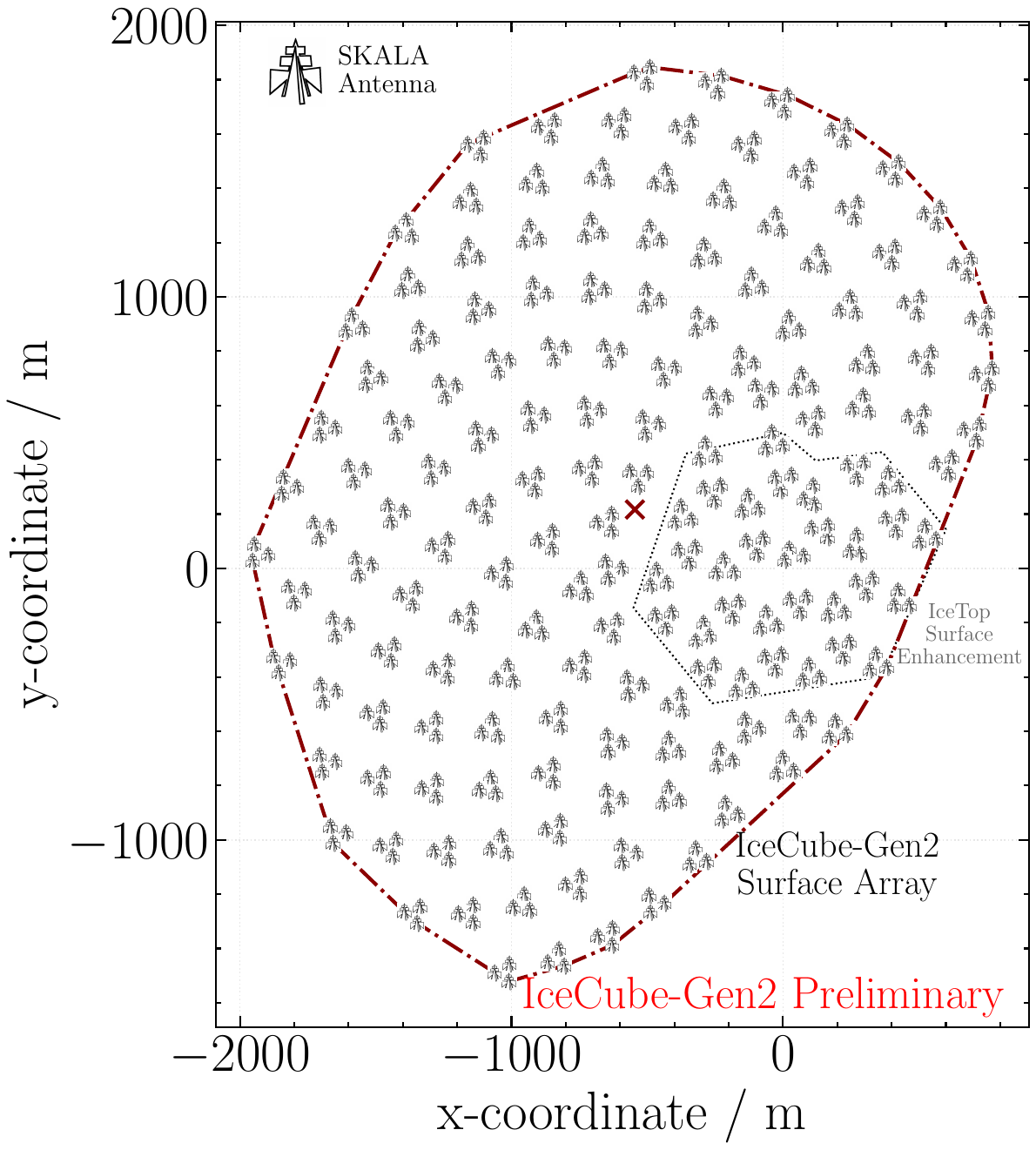}
    \caption{Proposed geometry of the Surface Array’s radio component for IceCube-Gen2, featuring SKALA antennas. The dashed-dotted (-$\cdot$-) red-line represents the boundary of the Surface Array, and red cross ($\times$) represents the geometric center of the array. The denser-array on top of IceTop is referred to as Surface Enhancement.}
    \label{fig:gen2Array}
\end{wrapfigure}
Hence, utilizing DNNs not only prevents the computational bottleneck but also removes the reliance on other detectors for observable inference. \par Using the proposed radio-antenna geometry of the planned IceCube-Gen2 Surface Array \cite{Aartsen2021} as a case study, this work develops a reconstruction pipeline to estimate the location of the air shower core, the direction of arrival, the primary energy, and $X_{\text{max}}$. The proposed array geometry is shown in \autoref{fig:gen2Array}. For the purpose of this analysis, we use simulations produced using CoREAS \cite{coreas}, in the energy range $8.5 \leq \mathrm{Log}_{10}(\mathrm{Energy/GeV}) \leq 9.4$. The simulation library consists of an almost equal range of the different primary nuclei (p, He, O, and Fe), in a zenith range between 0-75 degrees. To increase statistics for training, each CoREAS simulation is resampled ten times and thrown at a random location within 2500 m from the center of the Surface Array (denoted as $\times$ in \autoref{fig:gen2Array}). The antenna response is obtained by utilizing a star-pattern interpolation method, detailed in \cite{Abbasi2022}. Each antenna has two perpendicular polarizations, which are simulated independently.  For the majority of subsequent reconstructions, we utilize waveforms simulated at various antenna locations (and the corresponding polarizations). Modeled thermal and galactic noise (details in \cite{Abbasi2022}) is also added to the waveforms. The analysis further applies a cut on signal-to-noise ratio (SNR) to be above 100 (motivated by results in \cite{Schrder2024}), and the minimum number of triggered antennas after SNR-cleaning should be at least 9. The remaining simulation set is then used for further analysis.\par
\begin{wrapfigure}{r}{0.3\textwidth}
    \centering
    \includegraphics[width=\linewidth]{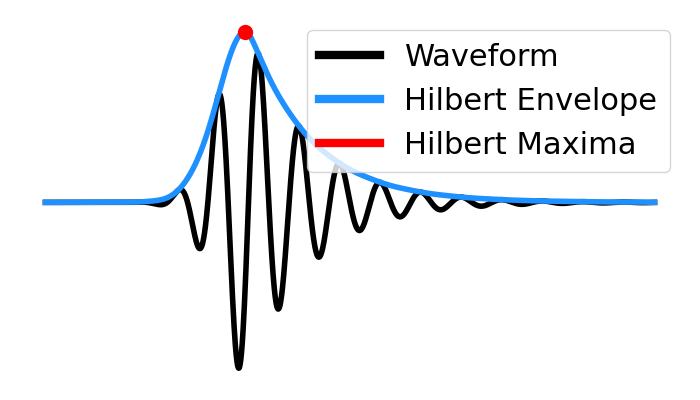}
    \caption{An example waveform with the Hilbert envelope and maxima marked.}
    \label{fig:examplewaveform}
\end{wrapfigure}
\begin{figure}[h]
\centering
\includegraphics[width=.85\linewidth]{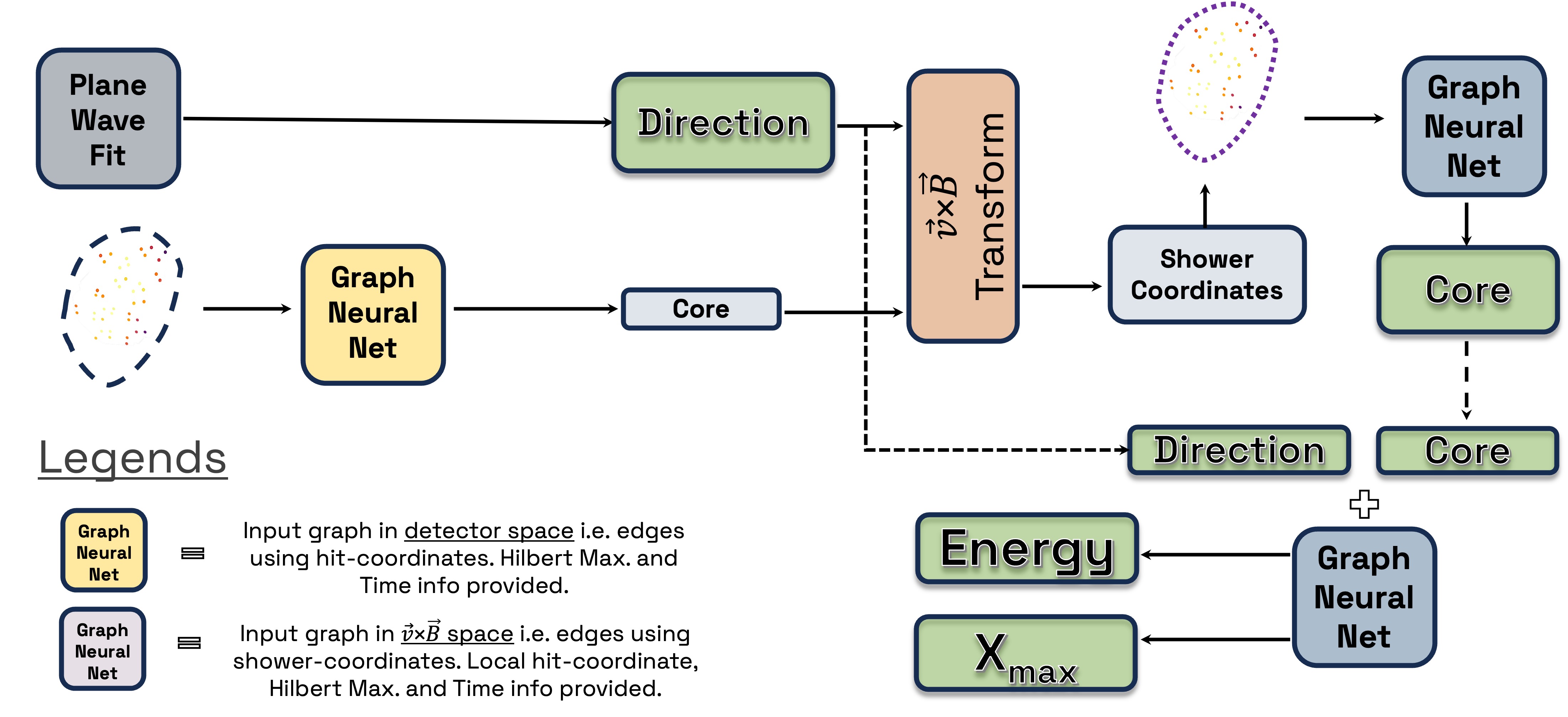}
\caption{
Schematic of the reconstruction pipeline combining physics-based and GNN-based methods. A plane-wave fit estimates the shower direction, while GNNs applied to detector-space and shower-coordinate graphs successively refine the core position. A final GNN predicts energy and \Xmax. Read \autoref{sec:method} for details and \autoref{sec:performance} for results.}
\label{fig:Pipeline}
\end{figure}

An artistic-expression of a radio waveform mentioned earlier is shown in \autoref{fig:examplewaveform} (black curve). Each waveform can be reduced to a single representative value using the maximum instantaneous amplitude called as Hilbert Maximum (HM), which is derived from the Hilbert envelope. The Hilbert envelope provides the instantaneous amplitude of the waveform, allowing a concise characterization through its maximum value. In addition to the amplitude of the HM, we will
 also utilize the time corresponding to the HM. The time is measured with respect to the time measured at the antenna closest to the HM-amplitude weighted centroid of the triggered antennas. The HM-amplitude and HM-time will be used multiple times for the rest of the analysis. The pipeline used for the various reconstructions is presented in \autoref{fig:Pipeline}. The pipeline utilizes both physics-based reconstructions and Graph Neural Network (GNN) based reconstructions. The multiple decisions made in the pipeline allow for fast and accurate reconstructions. The pipeline in multiple places uses GNNs, which are trained on graph-structured data. A graph is built of nodes and edges. For simplification, nodes can be interpreted as input pixels, and edges define the neighborhood between the nodes/pixels. 
The choice of GNNs over traditional CNNs was motivated by the irregular geometry of the proposed Surface Array, as well as the intention for this work to serve as a test case for other detector configurations with diverse geometries. Given their inherent flexibility and ability to operate on non-Euclidean domains, GNNs are the most natural architecture for this application. The following text will elaborate on each part of the pipeline.\par
\subsection{Direction Reconstruction}\label{subsec:direReco}
To infer the arrival direction of the cosmic-ray primaries using an antenna array, we assume that the wavefront can be approximated as a plane wave. By fitting a plane wave to the arrival times (HM-time) of the radio pulses at multiple antenna positions, one can estimate the zenith and azimuth angles of the incoming radiation. Although improved methods are available, the plane-wave method has been tested in other radio arrays such as LOFAR \cite{ Schellart2013, CORSTANJE201522} and was able to obtain an angular resolution of about 1 degree. 
The plane-wave reconstruction first includes the evaluation of an estimated core location. This is done by computing an HM-amplitude weighted centroid over triggered antenna positions. To obtain the estimated direction, the method utilizes HM-time of a specific polarization at each antenna. A plane perpendicular to the shower-axis, centered at the estimated core, is considered, and the method then tries to fit the direction (zenith and azimuth) that best fits the expected time with the HM-time (i.e., measured time). The results obtained from plane-wave fit reconstruction for direction estimate will be discussed in \autoref{subsec:perfDir}. Given that the current analysis remains at the proof-of-concept stage, this method is retained due to its simplicity and computational efficiency. 
\subsection{Core Reconstruction}\label{subsec:coreReco}
\begin{wrapfigure}{r}{0.5\textwidth}
    \centering
    \includegraphics[width=\linewidth]{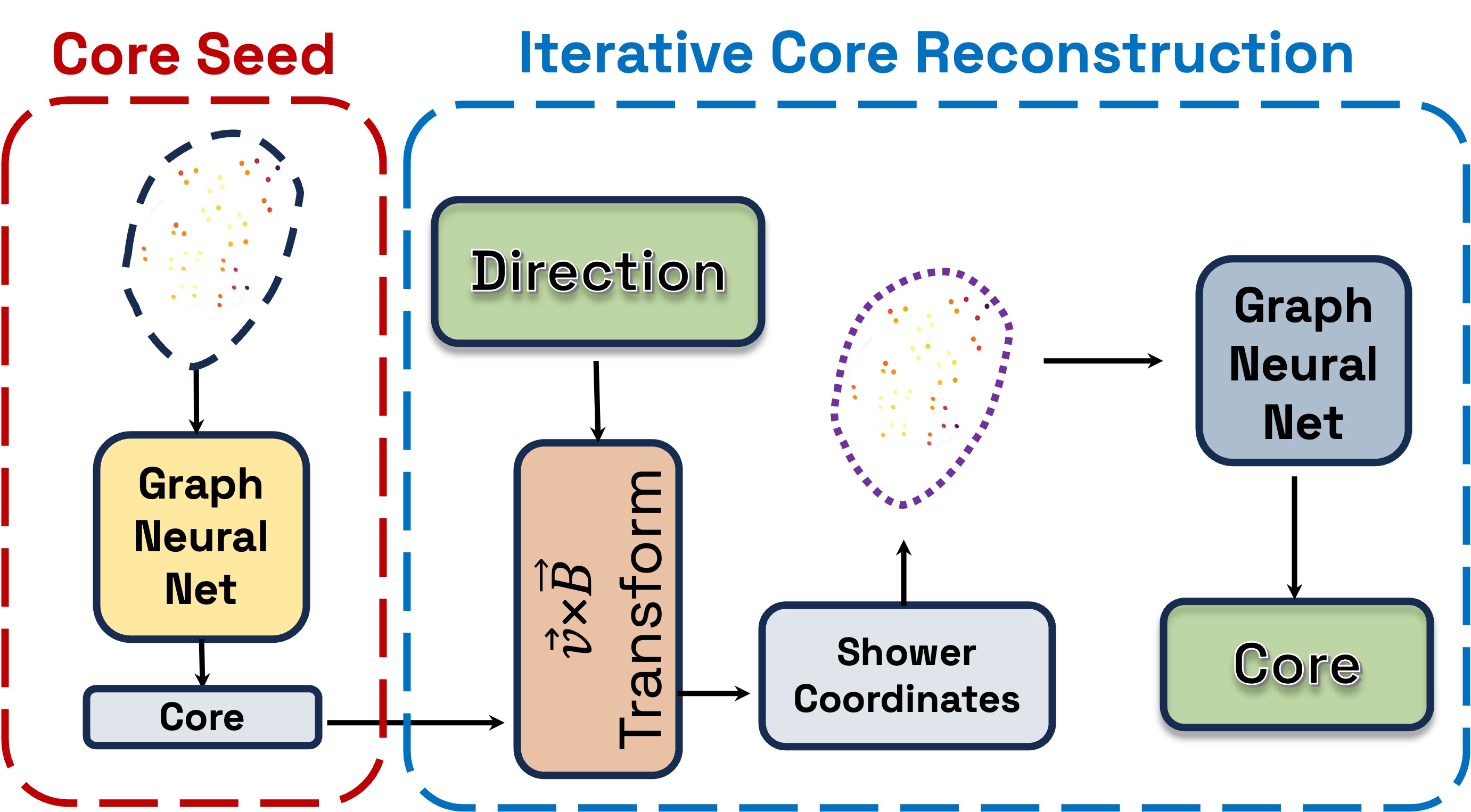}
    \caption{Part of pipeline from \autoref{fig:Pipeline} used for core estimate.}
    \label{fig:corePipeline}
\end{wrapfigure}

Even though the core estimate obtained by plane-wave reconstruction discussed in \autoref{subsec:direReco} is a good seed to perform direction reconstruction, the method by construction only allows for core reconstruction within the Surface-Array geometry, i.e., contained events.  Furthermore, since we know that the lateral distribution of the radio signal on the ground is azimuthally asymmetric \cite{Coleman2023}, a simple centroid estimation will be inherently biased. To provide a good core reconstruction, this work utilizes an iterative approach utilizing GNNs. The pipeline used for the core estimate using GNNs is shown in \autoref{fig:corePipeline}. The architecture of the GNN used in 
\autoref{fig:corePipeline} (as well as the upcoming ones) is motivated by \cite{Li2019}. Only minor changes in hyperparameter choice (e.g., number of layers, number of hidden-nodes, normalization-layer choice, etc) have been done. The reconstruction is done in two steps. The first step (referred to as "Core Seed" in \autoref{fig:corePipeline}) utilizes the triggered antennas to build a graph, i.e., antennas serve as nodes of the graph. Each node is connected to its 48 nearest neighbors, where the neighbors are evaluated by using the coordinate (x, y) of the antennas in the IceCube coordinate system. The amplitudes and times corresponding to HM for each polarization in an antenna waveform and the (x,y) coordinates of the antenna serve as features associated with each node. The network is trained to predict core location, with the true x and y coordinates from the simulation serving as the target variables. This is represented as "Core" in the "Core Seed" section of \autoref{fig:corePipeline}. This step provides an improved core reconstruction (for both contained and un-contained EAS) than plane wave reconstruction. \par
For vertical showers, the radio footprint in detector-coordinates generally looks circular. However, for more inclined showers the footprint of hits stretches out, and the footprint outline might appear as a thin ellipse \cite{SCHRODER20171}. Prior analysis experience from other observatories \cite{LofarXmax} has taught us that it is easier to reconstruct in shower-coordinates. \vxb serves as the x-axis in this coordinate system, where $\vec{v}$ is the direction of the shower axis, and $\vec{B}$ is the direction of the geomagnetic field at the detector location. Naturally, \vxvxb serves as the second-axis, perpendicular to \vxb. Hence, the next step in the pipeline is to transform the detector-hits coordinates into shower-coordinates. The prior core-estimate from the GNN ("Core Seed") and the direction estimate from \autoref{subsec:direReco} are used to perform this transformation and are represented as "\vxb Transform" in \autoref{fig:Pipeline}. The obtained coordinates are now utilized to build the edges of the graph in \vxb (or shower-coordinate) space. Additionally, the shower-coordinates are used as additional features for each of the graph nodes. A new GNN is now trained again to predict the core location where the difference from the "Core Seed" step is in graph edge construction, and shower coordinates are an additional input at each node. The transform and the GNN prediction step are labeled as "Iterative Core Reconstruction" in \autoref{fig:corePipeline}. Currently, this step is only performed once. However, it can be performed iteratively multiple times, with core prediction from each time serving as a seed for the \vxb transform. The final core-reconstruction performance will be discussed in \autoref{subsec:perfCore}.          

\subsection{Energy and X$_{\textbf{max}}$ Reconstruction}\label{subsec:energyAndXmaxReco}
Similar to the last step of core reconstruction, this step also trains a GNN where the graph is built in \vxb space and the node features include detector-coordinates, shower-coordinates, HM amplitude, and time (for both polarizations). In addition to this, the reconstructed direction (from \autoref{subsec:direReco}) and core (from \autoref{subsec:coreReco}) are also given as additional high-level inputs to the network. The network is trained to predict energy and distance to \Xmax. The distance to \Xmax is then used to calculate \Xmax. The combined prediction allows the network to also learn and resolve known degeneracies in footprint, which exist for different combinations of the two targets. The performance of energy and \Xmax reconstruction will be discussed in \autoref{subsec:perfEandXmax}.

\section{Performance}\label{sec:performance}
The following text details the reconstruction accuracy of core position, arrival direction, energy, and \Xmax using the pipeline illustrated in \autoref{fig:Pipeline}.
\subsection{Direction}\label{subsec:perfDir}
As discussed in \autoref{subsec:direReco}, plane-wave reconstruction is utilized to estimate the direction of the incident cosmic-ray primary. The reconstruction performance is shown as opening angle ($\Psi$) in \autoref{fig:dirReco}. Opening angle is the angular separation between the predicted and true directions, defined by their respective zenith and azimuth angles for all and contained events. Contained events are a subset of all events defined as events whose true EAS core lies within the surface array footprint. The median opening angle between the reconstructed and true directions remains below 1 degree across the full energy range. For the subset, i.e., contained events, the reconstruction accuracy improves further, indicating enhanced angular resolution due to a larger fraction of radio-emission being detected at the antennas than otherwise. The worsening performance with increasing energy is likely due to of decreasing atmospheric depth between the observation level and \Xmax, consistent with expectations \cite{Glaser2016}. Reconstruction performance can potentially be improved by using improved estimates of the shape of the radio wavefront \cite{Apel2014, Coleman2023} or GNN-based reconstructions (similar to the ones used in \autoref{subsec:coreReco} and \autoref{subsec:energyAndXmaxReco}).
\subsection{Core}\label{subsec:perfCore}
\begin{figure}[t]
    \centering
    \begin{subfigure}[t]{0.5\textwidth}
        \centering
        \includegraphics[width=\linewidth]{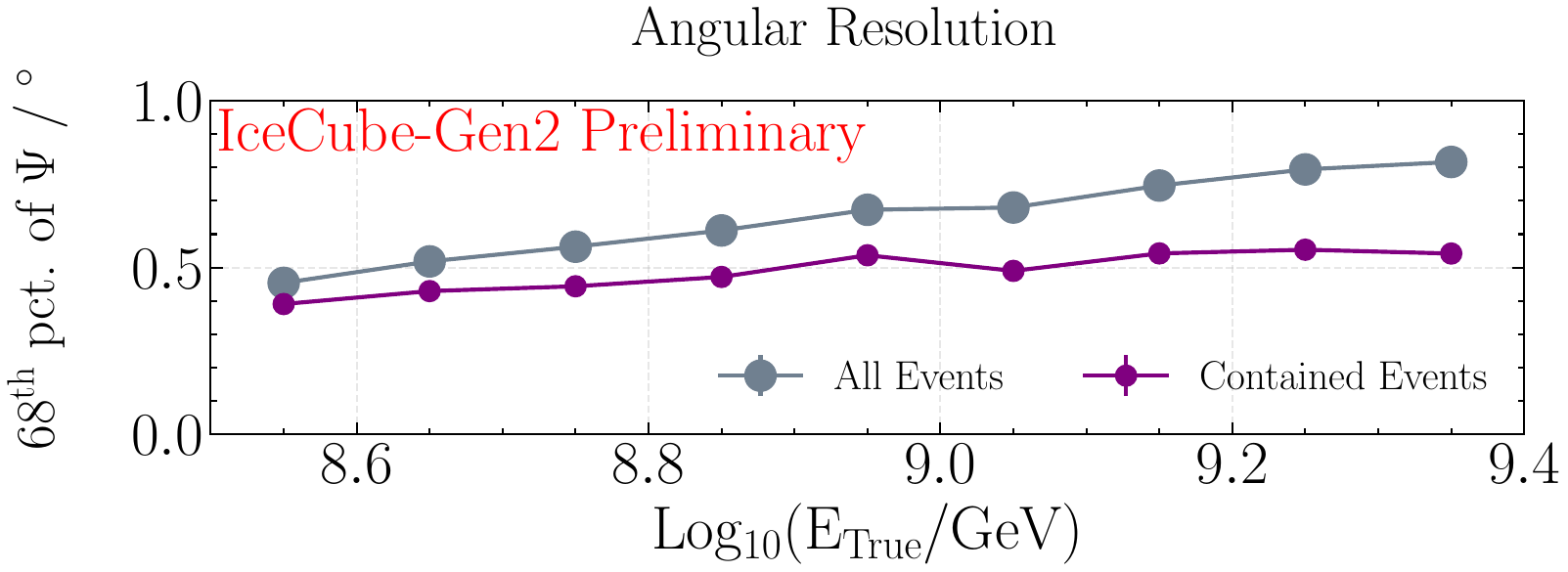}
        \caption{Direction Reconstruction}
        \label{fig:dirReco}
    \end{subfigure}%
    \hfill
    \begin{subfigure}[t]{0.5\textwidth}
        \centering
        \includegraphics[width=1\linewidth]{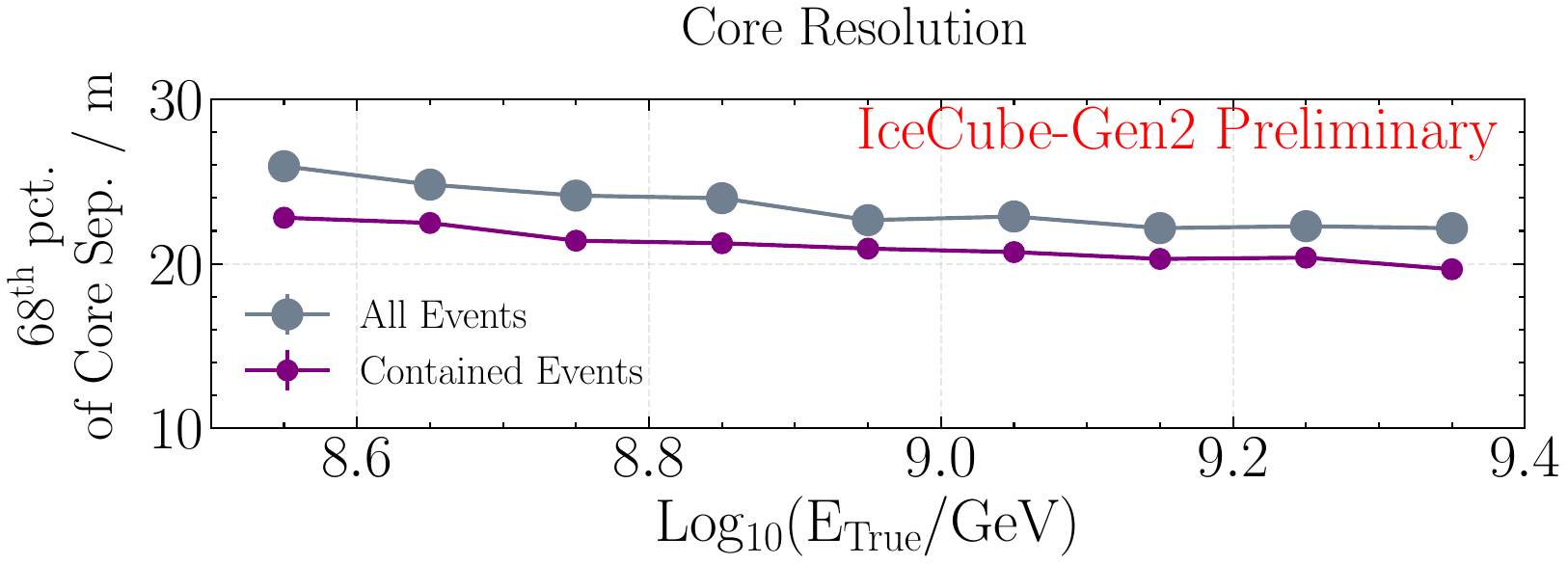}
        \caption{Core Reconstruction}
        \label{fig:coreReco}
    \end{subfigure}
    \caption{Performance of direction (left) obtained using plane-wave reconstruction (Section~\ref{subsec:perfDir}) and core (right) reconstruction using an iterative Graph Neural Network (Section~\ref{subsec:perfCore}).}
    \label{fig:RecoPerformance}
\end{figure}

As discussed in \autoref{subsec:coreReco}, core reconstruction is performed using an iterative approach based on Graph Neural Networks (GNNs). The reconstruction performance, shown in \autoref{fig:coreReco}, is quantified by the distance between the reconstructed and the true EAS core positions. The panel presents core resolution (68$^{th}$ percentile of core separations) as a function of energy for all events as well as contained events. Contained events have better resolution across all energies, as expected. The core resolution remains relatively stable across the energy range, with a weak trend toward improved accuracy at higher energies. The work already presents the robustness of the GNN-based method for a variety of EAS geometries as well as its effectiveness in reconstructing both contained and uncontained events.

\subsection{Energy and X$_{\textbf{max}}$}\label{subsec:perfEandXmax}
\begin{figure}[t]
    \centering
    \begin{subfigure}[t]{0.48\textwidth}
        \centering
        \includegraphics[width=\linewidth]{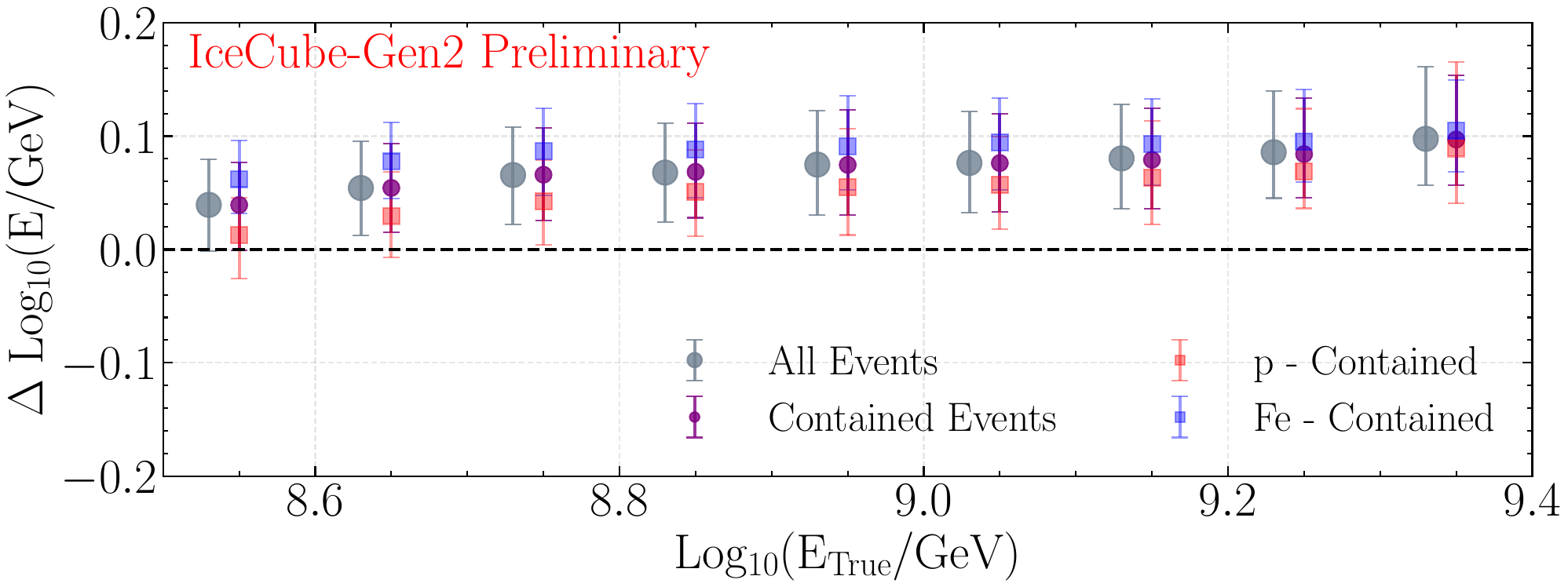}
        \caption{ Bias (median) and resolution (68$\%$ confidence interval) for energy prediction as a function of true energy.}
        \label{fig:energyReco}
    \end{subfigure}%
    \hfill
    \begin{subfigure}[t]{0.5\textwidth}
        \centering
        \includegraphics[width=\linewidth]{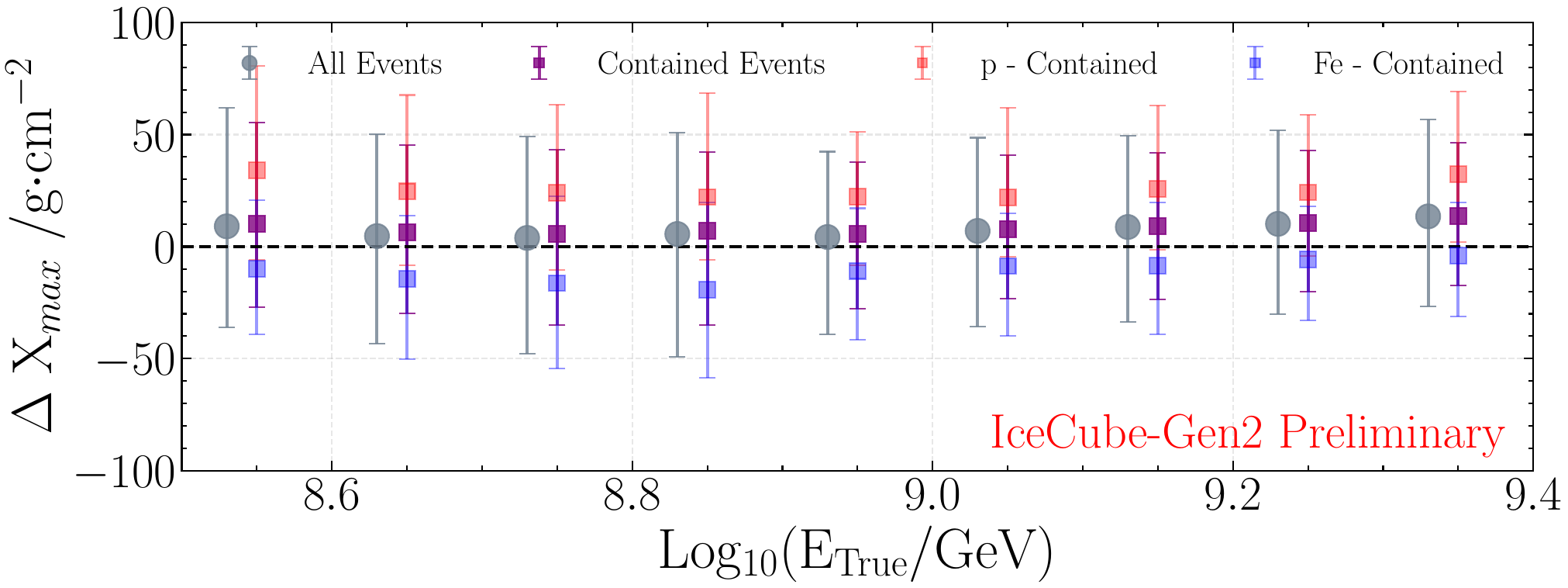}
        \caption{Bias and resolution for \Xmax prediction as a function of true energy.}
        \label{fig:xReco}
    \end{subfigure}
    \caption{Performance for energy (left) and \Xmax (right) reconstruction using Graph Neural Network. A shift in x-axis for "All Events" is introduced for easier visibility.}
    \label{fig:EnergyAndXmaxRecoPerformance}
\end{figure}

Similar to core prediction, energy and \Xmax reconstruction are also performed using GNNs. The reconstruction performance is shown in \autoref{fig:EnergyAndXmaxRecoPerformance}, as a function of primary energy. The energy reconstruction shows a bias for all EASs as well as for contained. Ongoing work is trying to reduce this bias.  The error bars (also for \autoref{fig:xReco})  indicate the 68 $\%$ confidence interval, and hence a measure of resolution. The figure also shows the energy bias and resolution for p and Fe.  The bias and resolution performance for \Xmax prediction is shown in \autoref{fig:xReco}. The prediction shows small overall bias for all as well as contained EAS. Although it still has primary-type dependent bias. The resolution is of the order 50 g$\cdot$cm$^{-2}$. Ongoing work is focusing on reducing the mass bias and improving the resolution. 
\section{Summary and Outlook}\label{sec:outllok}
This work presents a pipeline that integrates physics‐based reconstructions with Graph Neural Network reconstructions to reconstruct direction, core-location, energy, and \Xmax, using only radio‐antenna measurements. In the process, it reduces reliance on non-radio detectors and prevents the computational bottleneck of generating simulations for each detected event, as required by conventional reconstruction methods. Further improvement in energy and \Xmax reconstruction is needed. Overall, the work lays a foundation and demonstrate a new method to reconstruct both direction, core-location, energy, and \Xmax in a simulation-efficient manner. The pipeline can also be adapted to be used by hybrid observatories where reconstructions from other detectors can also be utilized or improved upon. A future development of this work also intends to test the reconstruction performance by adding measured noise and implementing denoising methods developed in \cite{Schrder2024}.

\bibliographystyle{ICRC}
{
\bibliography{references}}

\clearpage
\section*{Full Author List: IceCube-Gen2 Collaboration}

\scriptsize
\noindent
R. Abbasi$^{16}$,
M. Ackermann$^{76}$,
J. Adams$^{21}$,
S. K. Agarwalla$^{46,\: {\rm a}}$,
J. A. Aguilar$^{10}$,
M. Ahlers$^{25}$,
J.M. Alameddine$^{26}$,
S. Ali$^{39}$,
N. M. Amin$^{52}$,
K. Andeen$^{49}$
% ...
\\
\\
% ...
$^{10}$ Universit{\'e} Libre de Bruxelles, Science Faculty CP230, B-1050 Brussels, Belgium \\
% ...
$^{16}$ Department of Physics, Loyola University Chicago, Chicago, IL 60660, USA \\
% ...
$^{21}$ Dept. of Physics and Astronomy, University of Canterbury, Private Bag 4800, Christchurch, New Zealand \\
% ...
$^{25}$ Niels Bohr Institute, University of Copenhagen, DK-2100 Copenhagen, Denmark \\
$^{26}$ Dept. of Physics, TU Dortmund University, D-44221 Dortmund, Germany \\
% ...
$^{39}$ Dept. of Physics and Astronomy, University of Kansas, Lawrence, KS 66045, USA \\
% ...
$^{46}$ Dept. of Physics and Wisconsin IceCube Particle Astrophysics Center, University of Wisconsin{\textemdash}Madison, Madison, WI 53706, USA \\
% ...
$^{49}$ Department of Physics, Marquette University, Milwaukee, WI 53201, USA \\
% ...
$^{52}$ Bartol Research Institute and Dept. of Physics and Astronomy, University of Delaware, Newark, DE 19716, USA \\
% ...
$^{76}$ Deutsches Elektronen-Synchrotron DESY, Platanenallee 6, D-15738 Zeuthen, Germany \\
$^{\rm a}$ also at Institute of Physics, Sachivalaya Marg, Sainik School Post, Bhubaneswar 751005, India \\
% ...

\subsection*{Acknowledgments}

\noindent
The authors gratefully acknowledge the support from the following agencies and institutions:
USA {\textendash} U.S. National Science Foundation-Office of Polar Programs,
U.S. National Science Foundation-Physics Division,
U.S. National Science Foundation-EPSCoR,
U.S. National Science Foundation-Office of Advanced Cyberinfrastructure,
Wisconsin Alumni Research Foundation,
Center for High Throughput Computing (CHTC) at the University of Wisconsin{\textendash}Madison,
Open Science Grid (OSG),
Partnership to Advance Throughput Computing (PATh),
Advanced Cyberinfrastructure Coordination Ecosystem: Services {\&} Support (ACCESS),
Frontera and Ranch computing project at the Texas Advanced Computing Center,
U.S. Department of Energy-National Energy Research Scientific Computing Center,
Particle astrophysics research computing center at the University of Maryland,
Institute for Cyber-Enabled Research at Michigan State University,
Astroparticle physics computational facility at Marquette University,
NVIDIA Corporation,
and Google Cloud Platform;
Belgium {\textendash} Funds for Scientific Research (FRS-FNRS and FWO),
FWO Odysseus and Big Science programmes,
and Belgian Federal Science Policy Office (Belspo);
Germany {\textendash} Bundesministerium f{\"u}r Forschung, Technologie und Raumfahrt (BMFTR),
Deutsche Forschungsgemeinschaft (DFG),
Helmholtz Alliance for Astroparticle Physics (HAP),
Initiative and Networking Fund of the Helmholtz Association,
Deutsches Elektronen Synchrotron (DESY),
and High Performance Computing cluster of the RWTH Aachen;
Sweden {\textendash} Swedish Research Council,
Swedish Polar Research Secretariat,
Swedish National Infrastructure for Computing (SNIC),
and Knut and Alice Wallenberg Foundation;
European Union {\textendash} EGI Advanced Computing for research;
Australia {\textendash} Australian Research Council;
Canada {\textendash} Natural Sciences and Engineering Research Council of Canada,
Calcul Qu{\'e}bec, Compute Ontario, Canada Foundation for Innovation, WestGrid, and Digital Research Alliance of Canada;
Denmark {\textendash} Villum Fonden, Carlsberg Foundation, and European Commission;
New Zealand {\textendash} Marsden Fund;
Japan {\textendash} Japan Society for Promotion of Science (JSPS)
and Institute for Global Prominent Research (IGPR) of Chiba University;
Korea {\textendash} National Research Foundation of Korea (NRF);
Switzerland {\textendash} Swiss National Science Foundation (SNSF).

\end{document}